\documentclass[doublecol,a4paper,showpacs]{epl2}
\usepackage{graphicx}
\usepackage[all]{xy}
\usepackage{amsmath}
\usepackage{amssymb}
\usepackage{tensor}
\newcommand{\be}{\begin{equation}}
\newcommand{\ee}{\end{equation}}
\newcommand{\ben}{\begin{eqnarray}}
\newcommand{\een}{\end{eqnarray}}
\newcommand{\bes}{\begin{subequations}}
\newcommand{\ees}{\end{subequations}}
\def\bal#1\eal{\begin{align}#1\end{align}}

\newcommand{\bb}{\bibitem}

\begin{document}
\title{Complete factorization of equations of motion for generalized scalar field theories}

\author{D. Bazeia\and Diego R. Granado\thanks{{corresponding author;\email{\,diegorochagranado@gmail.com}}}\and Elisama E.M. Lima{\shortauthor{D. Bazeia \etal}}}

\institute{{Departamento de F\'\i sica, Universidade Federal da Para\'\i ba, 58051-970 Jo\~ao Pessoa, Para\'\i ba, Brazil}}

\date{\today}
\abstract{We demonstrate that the complete factorization of equations of motion into first-order differential equations can be obtained for real and complex scalar field theories with non-canonical dynamics.}


\maketitle

\section{Introduction}

In 1934, Born and Infeld (BI) proposed a nonlinear generalization of electrodynamics \cite{Born:1934gh} to avoid the divergence problem with the electron self-energy. It is an important example where nonlinear dynamics may play a role and it was later shown that such action, now called DBI action \cite{Born:1934gh,Dirac}, arises naturally in the context of string theory \cite{stringdbi}. Some examples of its usage can be seen as, for instance: an effective action for the tachyon condensation \cite{tachyonstring}; a string theory model to describe the early inflationary period of the universe \cite{dbiinflation}; in cosmology a scalar DBI action can be used to describe the acceleration of the universe \cite{kessence}, as well as the inflationary period in the early universe \cite{kinflation}; models with the presence of instanton solutions \cite{Brown}, the description of global strings \cite{Sarangi}; vortex solutions \cite{Babichev1,VortexBI}; formulation of twinlike models \cite{Andrews} and so on. More recently, several analytical models described within the DBI context admitting kinks solutions were explored in Ref.~\cite{Lima}. The DBI model opened up a new road to consider
non-canonical dynamics and to deal with more general forms of the kinetic term. 

Domain walls and branes have drawn a lot of attention to the physics community \cite{vafa} and may appear when the scalar field has nonlinear dynamics as well. Some examples include: in \cite{Hora} one investigates specific features of kinks and vortices; in \cite{Adam,FOFGD,CompGD,BLMM} one studies how the modifications on the kinetic part of the Lagrangian introduce new nonlinear terms to the equations of motion, which can permit the appearance of compact solutions \cite{Adam,FOFGD,CompGD}, also known as compactons \cite{Rosenau}; in \cite{SPGD} the defect structure within a scalar generalized profile is studied; and in \cite{Olechowski,BGD} the braneworld scenario is also investigated.  

In \cite{CFO1,CFO2} the authors worked out the first order formalism in Wess-Zumino theory and in the bosonic sector of the supersymmetric field theory. In both papers, by means of the canonical field theory, the authors showed that not just the solutions of the first order equation solves the second-order equation of motion but the reverse path is also valid, \textit{i.e.}, by means of the second-order equation of motion we can reach the first-order BPS equations, under certain circumstances.

In this paper we work with real and complex scalar field, and consider the case of a non-canonical dynamics. The goal is to generalize the method proposed in \cite{CFO1,CFO2} to a wider class of scalar field dynamics, \textit{i.e.}, to find the first order differential equation through the second order equation of motion in a general kinematic setup. To do this, we start dealing with the general framework for real and then for complex scalar field theories, and then end the work including our comments and conclusions. 

\section{The Framework}
\label{sec-1}

Let us start with a single real or complex scalar field in $(1,1)$ spacetime dimensions, with metric such that $x^\mu=(t,x)$. We also use natural units, with $\hbar=c=1$. In the case of standard dynamics, the models are given by, for $\phi$ being a real scalar field,
\be 
{\cal L}=\frac12\partial_\mu\phi\partial ^\mu\phi-V(\phi),
\ee
and by, for $\varphi$ being a complex field,
\be 
{\cal L}=\partial_\mu{\bar\varphi}\partial^\mu\varphi-V(|\varphi|).
\ee
We write, for simplicity,
\be  
X=\frac12\partial_\mu\phi\partial^\mu\phi
\ee
in the case of a real field, and
\be  
X=\partial_\mu{\bar\varphi}\partial^\mu\varphi,
\ee
in the case of a complex field. So, we can write the standard Lagrangian as
\be  
{\cal L}=X-V.
\ee

In this section, we want to study scalar fields with non-standard dynamics, so we write
\be\label{gmodel}
{\cal L}=F(X)-V,
\ee
where $F(X)$ is a function of $X$, which can have several distinct forms; for instance, in the case of a DBI modification it can be written in the form
{{\be  
F(X)=a-a\sqrt{1-\frac2{a}\;X\,},
\ee}}
such that for $a$ very large it reproduces the standard dynamics, up to first order in $(1/a)$; see, e.g., the  recent work \cite{Lima}. Another possibility that may lead to the generation of compact structures could be controlled by the generalized dynamics \cite{FOFGD,BLMM}
\be
F(X)=-X^2.
\ee
However, as we are dealing with non-canonical dynamics one leaves $F(X)$ as generic as possible.

\subsection{Real field}
\label{sub-1}

Let us focus on the real scalar field model.
The Lagrangian \eqref{gmodel} gives rise to the energy-momentum tensor
\be
T_{\mu\nu}=F_{X}\partial_{\mu}\phi\partial_{\nu}\phi-\eta_{\mu\nu}{\cal L},
\ee
where $F_{X} = dF/dX$. The equation of motion for the scalar field has the form
\be
\label{eom0}
\partial_{\mu}(F_{X}\partial^{\mu}\phi)+V_{\phi}=0.
\ee
As we are interested in defect structures, we assume static configurations, $\phi = \phi(x)$; here we get
\be
\label{eom1}
(F_X+2XF_{XX})\phi''=V_{\phi},
\ee
where the prime denotes derivative with respect to $x$, and now $X = -\phi'^2/2$.  Multiplying the equation \eqref{eom1} by $\phi'$ and integrating it, we obtain the following first-order equation
\be
\label{fore}
F-V-2XF_{X}=0,
\ee
where the stressless condition was applied, that is, we used $T_{11} = 0$. 

The energy density is $\rho(x) = T_{00} =-F +V$, and from the equation \eqref{fore} it becomes
\be
\label{rho}
\rho(x)=F_{X}\phi'^2.
\ee
By means of the formalism introduced in Ref.~\cite{FOFGD}, we suppose the existence of another function of the scalar field  $W = W(\phi)$, such that
\be
\label{nfoe}
F_X \phi'=W_{\phi},
\ee
where $W_\phi=dW/d\phi$. This allows us to write the energy density in the form $\rho(x) = W_{\phi} \phi' = dW/dx$, so that the energy becomes
\be
E=W(\phi(x\rightarrow\infty))-W(\phi(x\rightarrow-\infty)).
\ee
The first-order equation \eqref{fore} can also be written as
\be
\label{fpot}
W_{\phi}\phi'=-{\cal L}.
\ee
From \eqref{nfoe} we have that the necessary condition for the field $\phi(x)$ to satisfy the equation of motion \eqref{eom1} is
\be
\label{ssoe}
V_{\phi}=W_{\phi\phi}\phi'.
\ee
If Eq.~\eqref{fpot} combined with \eqref{nfoe} agrees with this condition, then one can say that these equations provide solutions for the equation of motion. 

The above requirement must be tested for each one of the chosen dynamics. In particular, for $F(X) = X$, these equations result in the usual first-order description of BPS states: $\phi'=W_{\phi}$ and $V(\phi)=\frac{1}{2}W_{\phi}^2$  \cite{BPS}. 

Until now, we have suggested  the possibility of factorizing the equation of motion of a real scalar field with a general kinetic term  into  first-order equations,  in a way compatible with the BPS formalism \cite{BPS}. In order to demonstrate the equivalence between the second-order equation and the  first-order equations adopted here, as in \cite{CFO2}, we define the quantity $R(\phi)$ in the form
\be
R(\phi)=\frac{F_{X}\phi'}{W_{\phi}}.
\ee 
At this point we would like to remark that in the canonical standard case we have $F_X=1$. Thus we recover the ratio definition proposed in \cite{CFO2}. From the definition above we can write
\begin{eqnarray}
\frac{d R(\phi)}{dx}&=&\left(W_{\phi}\frac{d}{dx} (F_{X}\phi')-F_{X}\phi'^2W_{\phi\phi}\right)\frac{1}{W_{\phi}^2},\nonumber\\
&=&\left(W_{\phi}V_{\phi}-F_{X}\phi'^2W_{\phi\phi}\right)\frac{1}{W_{\phi}^2},\nonumber\\
&=&\left(W_{\phi}^2-(F_{X}\phi')^2\right)\frac{W_{\phi\phi}}{F_X W_{\phi}^2},
\end{eqnarray}
after using the second-order equation of motion for static configurations, given by $ (F_{X}\phi')'=V_{\phi}$. Also, we have used that the potential satisfies the relation $F_X V_{\phi}=W_{\phi}W_{\phi\phi}${, obtained as a consequence of the equations \eqref{nfoe} and \eqref{ssoe}}.

Turning our attention to the following quantity 
\be
S(\phi)=W_{\phi}^2-(F_{X}\phi')^2,
\ee
one can see that 
\ben
\frac{d S(\phi)}{dx}&=&2W_{\phi}W_{\phi\phi}\phi'-2F_X\phi'\frac{d}{dx} (F_{X}\phi') \nonumber\\
&=& 2W_{\phi}W_{\phi\phi}\phi'-2F_X\phi'\left(\frac{W_{\phi}W_{\phi\phi}}{F_X}\right) \nonumber \\
&=&0. \nonumber
\een
Thus the quantity $S(\phi)$ is constant  with respect to $x$ when $\phi(x)$ is solution of the equation of motion. 

The model is supposed to support BPS solutions, so the static field must obey the boundary conditions $\lim_{x\rightarrow - \infty} \phi(x)= v_{k}$, where $v_{k}$ represents a minimum such that $\lim_{x\rightarrow - \infty} \phi'(x)= 0$. Also, we assume that the minima are extrema of the function $W(\phi)$, {\it{i.e}}.,  $\lim_{x\rightarrow - \infty} W_{\phi}(\phi(x))= 0$, thus
\be
\lim_{x\rightarrow - \infty} S(\phi)= \lim_{x\rightarrow - \infty} \left(W_{\phi}^2-(F_{X}\phi')^2\right)=0.
\ee
These statements show that $S(\phi)$ vanishes, then we get $R=1$ which provides the first-order equation $F_X \phi'= W_{\phi}$. This shows that we can find BPS solutions in a more general framework, that are equivalent to the solutions of the equation of motion. This result is a extension to the case of non-canonical dynamics of the case described by a single real scalar field with canonical dynamics, as described in Ref.~{\cite{CFO1,CFO2}}. 

\subsection{Complex Field} \label{sub-2}

Here we want to extend the generalized formalism of the previous subsection to the case of the complex scalar fields $\varphi$. Thus, we have
$X = \partial_{\mu}\varphi\partial^{\mu}\overline\varphi$ and $V(|\varphi|)$ for the real potential energy.
 
In this case the energy-momentum tensor reads:
\be
T_{\mu\nu}=F_{X}\left(\partial_{\mu}\varphi\partial_{\nu}\overline\varphi+\partial_{\mu}\overline\varphi\partial_{\nu}\varphi\right)-\eta_{\mu\nu}{\cal L}.
\ee
The equation of motion for the  field $\varphi(x, t)$ is
\be 
\partial_{\mu}(F_{X}\partial^{\mu}\varphi)+V_{\overline\varphi}=0,
\ee
where $V_{\overline\varphi}=\partial V/\partial{\overline{\varphi}}$. Remember that $V_{\overline{\varphi}}=(\varphi/2|\varphi|)\,V_{|\varphi|}$, $V_{\varphi}=(\overline{\varphi}/2|\varphi|)\,V_{|\varphi|}$, and $V_{\overline\varphi}=\overline{V_\varphi}$.
Since we are searching for defect structures, we consider static configurations. In this case the above equation of motion becomes
\be
\label{eom1c}
\frac{d}{dx}(F_{X}\varphi')=V_{\overline\varphi},\\
\ee
where now $X = -\varphi'\overline\varphi'$.

Following the steps presented in the previous subsection we get
\be
\label{forec}
F-V-2XF_{X}=0.
\ee
This equation \eqref{forec} obeys the stressless condition, and 
from it the energy density reads
\be
\label{rhoc}
\rho(x)=-2XF_{X}=2\varphi'\overline\varphi'F_{X}.
\ee
By considering $W = W(\varphi)$ as a holomorphic function we have
\be
\label{nfoec}
F_{X} \varphi'=\overline{W_{\varphi}} e^{-i\xi},
\ee
where $\xi$ is a real parameter. With this, the first-order equation \eqref{forec} may be written as
\be\label{fpotc}
-{\cal L}=2\varphi'\, W_{\varphi}e^{i\xi}.
\ee
Thus the energy density becomes
\begin{eqnarray}
 \rho(x) &=&2\varphi'W_{\varphi}e^{i\xi} \nonumber\\
&=&2\frac{d}{dx}\left(W(\varphi)e^{i\xi}\right) \nonumber\\
&=& 2\frac{d}{dx}\left|W(\varphi)\right|,
\end{eqnarray}
and the energy is $E=2|\Delta W|$, where
\be
\Delta W=\left|W\left(\varphi(x\rightarrow\infty)\right)\right|-\left|W\left(\varphi(x\rightarrow-\infty)\right)\right|.
\ee
 From \eqref{nfoec} the equations of motion \eqref{eom1c} are satisfied if one writes
\ben
\label{ssoec}
V_{\varphi}=W_{\varphi\varphi}\varphi' e^{i\xi}.
\een
This is a first-order differential equation which together with \eqref{nfoec} provides solutions for the equations of motion. In the standard canonical case we have $F(X)=X$. Thus, from Eqs.~\eqref{nfoec} one obtains  $\varphi'=\overline{W_{\varphi}} e^{-i\xi}$ and $ \overline\varphi '= W_{\varphi} e^{i\xi}$, whose solutions solve the equations of motion, since the Eq.~\eqref{fpotc} gives $V(\varphi,\overline\varphi)=\left|W_{\varphi}\right|^2$. This is the result found in \cite{CFO1,CFO2}.

The issue here is to extend the result of \cite{CFO1} to the case of non-canonical dynamics. With this in mind, we introduce the function $R(\varphi,\overline\varphi)$ defined as
\be
\label{Rcomp}
R(\varphi,\overline\varphi)=\frac{F_{X}\varphi'}{\overline{W_{\varphi}}} 
\ee
For $F_X=1$ we get to the canonical theory and recover the definition in \cite{CFO1}. From the new definition we have
\begin{eqnarray}
\frac{d R}{dx}&=&\left(\overline{W_{\varphi}}\frac{d}{dx} (F_{X}\varphi')-F_{X}\varphi'\overline{W_{\varphi\varphi}}\overline\varphi'\right)\frac{1}{\overline{W_{\varphi}}^2}
\nonumber\\
&=&\left(\left|W_{\varphi}\right|^2-\left|F_{X}\varphi'\right|^2\right)\frac{\overline{W_{\varphi\varphi}}}{F_X \overline{W_{\varphi}}^2}\nonumber\\
&=&\left(\left|W_{\varphi}\right|^2-\left|F_{X}\varphi'\right|^2\right)\frac{\overline{W_{\varphi\varphi}}}{F_X \overline{W_{\varphi}}^2}
\end{eqnarray}
where we have used the second-order equation for static fields, $ (F_{X}\varphi')'=\overline{V_{\varphi}}=W_{\varphi}\overline{W_{\varphi\varphi}}/F_X$. Defining the quantity $S(\varphi,\overline\varphi)=\left|W_{\varphi}\right|^2-\left|F_{X}\varphi'\right|^2$,  one can see that
\ben
\frac{d }{dx}\left|W_{\varphi}\right|^2&=&\frac{d }{dx}\left(W_{\varphi}\overline{W_{\varphi}}\right),\nonumber \\
&=& W_{\varphi\varphi}\overline{W_{\varphi}}\varphi'+W_{\varphi}\overline{W_{\varphi\varphi}}\overline\varphi', \nonumber
\een
and
\ben
\frac{d }{dx}\left|F_{X}\varphi'\right|^2&=&\frac{d }{dx}\left((F_{X}\varphi')(F_{X}\overline\varphi')\right),\nonumber \\
&=& F_{X}\overline\varphi'\frac{d}{dx} (F_{X}\varphi')+F_{X}\varphi'\frac{d}{dx} (F_{X}\overline\varphi'), \nonumber\\
&=&F_{X}\overline\varphi'\overline{V_{\varphi}}+F_{X}\varphi'V_{\varphi}, \nonumber\\
&=&W_{\varphi} \overline{W_{\varphi\varphi}}\overline\varphi'+\overline{W_{\varphi}}W_{\varphi\varphi}\varphi'. \nonumber
\een
Then, we get $dS(\varphi,\overline\varphi)/dx=0$; thus, $S(\varphi,\overline\varphi)$ is independent of $x$ when $\varphi(x)$ solves the equation of motion and
$F_XV_{\varphi}=\overline{W_{\varphi}}W_{\varphi\varphi}$. 

As pointed out before, in order to have BPS solutions we have $\lim_{x\rightarrow - \infty} \varphi(x)= v_{k}$, where $v_{k}$ is a minimum, and
$\lim_{x\rightarrow - \infty} \varphi'(x)= 0$. Again, one assumes that the minima are extrema of the holomorphic function $W(\varphi)$,
\textit{i.e.},  $\lim_{x\rightarrow - \infty} W_{\varphi}(\varphi(x))= 0$. Thus, we can write
\be
\lim_{x\rightarrow - \infty} S(\varphi,\overline\varphi)= \lim_{x\rightarrow - \infty} \left(\left|W_{\varphi}\right|^2-\left|F_{X}\varphi'\right|^2\right)=0. \nonumber \\
\ee
This means that $S(\varphi,\overline\varphi)$ vanishes, then  $\left|W_{\varphi}\right|^2=\left|F_{X}\varphi'\right|^2$ and from Eqs.~\eqref{Rcomp} one gets
\begin{equation}
\left|R(\varphi,\overline\varphi)\right|^2=R(\varphi,\overline\varphi)\overline{R}(\varphi,\overline\varphi)=\frac{1}{|W_{\varphi}|^2}|F_{X}\varphi'|^2=1. \nonumber \\
\end{equation}
The result gives $R= e^{-i\xi}$, which provides the first-order equations $F_X \varphi'=\overline{W_{\varphi}} e^{-i\xi}$. This demonstration extends the result obtained in \cite{CFO1,CFO2} to a complex field theory with non-canonical dynamics.

\section{Comments and conclusions}\label{sec-com}

In this work we investigated real and complex scalar fields in $(1,1)$ spacetime dimensions. In the two cases, the models of the scalar fields are controlled by non-canonical dynamics.

In Refs.~\cite{CFO1,CFO2} the authors proposed a method to show that through the equations of motion we can obtain the BPS first-order equations. They worked with standard scalar field theories. In the current work we showed that this method can be adapted to be used in a more general context, where the dynamics is controlled by non-canonical contributions. The results show that for the real scalar field the equation of motion is shown to be equivalent to the first-order equations. In the case of a complex field, the same is valid, but now one requires the presence of an holomorphic function of the complex field. This is also required in the case of standard dynamics, so our results extend the previous ones to this new scenario, showing the equivalence between the equations of motion and the first-order equations, despite the dynamics being canonical or non-canonical.
Since the motivation to use non-canonical dynamics has been increased in the recent years, we believe that the results of the work are also of current interest. 

{In the Ref. \cite{FOFGD} the authors proposed a general setup for the first-order formalism in a multi-component scalar fields with non-standard dynamics. Based on this setup, we are now investigating the possible extensions of the analysis implemented in this work to the case of multi-component scalar fields.} 

The authors would like to thank the Brazilian agency CNPq for financial support. 

\end{document}